\documentclass[twocolumn]{aastex702}

\newcommand{\cm}{cm$^{-1}$}

\newcommand{\NiII}{\ion{Ni}{2}}
\newcommand{\Nids}{3d\textsuperscript{8}4s}
\newcommand{\Nidp}{3d\textsuperscript{8}4p}
\newcommand{\Nidd}{3d\textsuperscript{8}4d}
\newcommand{\Nid}{3d\textsuperscript{9}}
\newcommand{\loggf}{$\log(\textsl{g}\!f)$}
\newcommand{\gf}{$\textsl{g}\!f$}
\newcommand{\pybr}{\texttt{PYBRANCH}}
\newcommand{\xgrem}{\texttt{Xgremlin}}
\usepackage{gensymb}
\usepackage{rotating}
\usepackage{multirow}
\usepackage{array}
\usepackage{tablefootnote}
\usepackage{nicefrac}
\usepackage{longtable}
\usepackage{hhline}
\usepackage{placeins}
\usepackage{siunitx}
\usepackage{threeparttable}
\usepackage{amsmath}
\newif\ifjournalshort
\journalshorttrue

\begin{document}

\title{Experimental {\NiII} Oscillator Strengths for Transitions from {\Nidd} to {\Nidp} Levels Measured Using High-resolution Fourier Transform Spectroscopy}

\author[0009-0000-9471-6242]{Ruijie Chen}
\affiliation{Department of Physics, Imperial College London, London, SW7 2AZ, UK}
\email{ruijie.chen21@imperial.ac.uk}

\correspondingauthor{Christian Clear}
\email{christian.clear@imperial.ac.uk}

\author[0000-0002-3339-7097]{Christian P. Clear}
\affiliation{Department of Physics, Imperial College London, London, SW7 2AZ, UK}
\email{christian.clear@imperial.ac.uk}

\author[0000-0002-1718-9650]{Gillian Nave}
\affiliation{Department of Physics, Imperial College London, London, SW7 2AZ, UK}
\affiliation{National Institute of Standards and Technology, Gaithersburg, MD, 20889-8422, USA}
\affiliation{NASA Goddard Space Flight Center, Greenbelt, MD, USA}
\email{g.nave@imperial.ac.uk}

\author[0000-0003-2879-4140]{Juliet C. Pickering}
\affiliation{Department of Physics, Imperial College London, London, SW7 2AZ, UK}
\email[]{j.pickering@imperial.ac.uk}



\begin{abstract}
We present the first experimentally measured oscillator strengths for {\NiII}  {\Nidd} -- {\Nidp}  transitions. High-resolution Fourier transform  spectra  were recorded of Ni-He hollow cathode lamps, and intensity calibrated using deuterium standard lamps to determine relative intensities of {\NiII} emission lines between 200 and 395 nm. Branching fractions were determined for observed lines from 35 {\Nidd} upper levels in total, and account for at least 95\% of the predicted transition probability for 31 of these levels, with sufficient completeness to give reliable branching fractions. Combining our measured branching fractions with four experimental and 27 theoretically calculated energy level lifetimes provided absolute oscillator strengths for 174 lines, for which no previous experimental values exist. We compare the results with three sets of previously published theoretical calculations, illustrating the improvement of laboratory measurements for these atomic data.
\end{abstract}


\section{Introduction}
Accurate atomic data are essential for the interpretation of high-resolution astrophysical spectra. Specifically, reliable oscillator strengths (transition probabilities, {\loggf}s) are required to generate realistic synthetic spectra and determine accurate elemental abundances. These yield valuable information about stellar and galactic chemical evolution and nucleosynthetic processes. 

The iron-group elements are of particular importance since their partially-filled 3d shells produce rich, dense spectra, and their relatively high cosmic abundances make them dominant contributors to stellar opacities.  {\NiII} has been observed in various astrophysical systems, including stars of different spectral types \citep{Jaschek1995}, supernovae \citep{Chen2024},  gamma ray bursts \citep{PUgliese2024}, and quasars \citep{Boisse2019}. Improved oscillator strengths for {\NiII} spanning a broad range of excitation potentials are therefore needed for modelling of these plasmas. Such atomic data must be as complete and accurate as possible since transitions suitable for abundance determinations vary with stellar metallicity and atmospheric conditions. When strong lines become saturated and unreliable for abundance determination, weaker transitions from higher-lying, less populated levels could remain optically thin. Chemical abundances derived from lines in different spectral regions also serve as diagnostics of 3D and non-local thermodynamic equilibrium (non-LTE) effects, enabling the development of new photospheric models that take these effects into account. Despite their importance to spectra of many astrophysical bodies, oscillator strengths currently available for {\NiII} are limited to transitions involving relatively low-lying levels only. This work on {\NiII} is part of a continuing effort to improve and extend experimental atomic data for singly-ionised species of the iron-group: \ion{Ti}{2} \citep{Pickering2001TiII}, \ion{V}{2} \citep{Thorne2013}, \ion{Fe}{2} \citep{Pickering2001FeII, Pickering2002, Johansson2002}, \ion{Co}{2} \citep{Ding2020}, \ion{Mn}{2} \citep{Liggins2021} and {\NiII} \citep{Clear2022, Clear2023, Clear2023[NiII]}.

Previous experimental studies of {\NiII} lifetimes, transition probabilities, and oscillator strengths (and {\loggf}s) have been performed with a variety of techniques, with a primary focus on the low-lying {\Nidp} -- {\Nids} and {\Nidp} -- {\Nid} transition arrays.  The earliest laboratory measurements of relative oscillator strengths for {\NiII} were obtained using the wall-stabilised arc method by \citet{Bell1966}, \citet{Heise1974}, \citet{Goly1975} and \citet{Moity1978}. Later studies combined experimental branching fractions (BFs) with level lifetimes to obtain absolute oscillator strengths \citep{Lawler1987,Fuhr1988,Ferrero1997, Fedchak1999}. For example, \citet{Fedchak1999} performed lifetime measurements with the time-resolved laser-induced fluorescence (TR-LIF) technique, which has become the standard practice, and reported lifetimes for 18 levels of the {\Nidp} configuration. These lifetimes were used to normalise branching fractions from Fourier transform (FT) and vacuum echelle spectra for 59 transitions. Among these were three {\NiII} resonance transitions involving the ground configuration {\Nid}, representing the first measurements of such transitions below 200 nm. Shortly after, six additional resonance lines were measured by \citet{Fedchak2000} using absorption spectroscopy. In addition, \citet{Manrique2011, Manrique2013} determined absolute oscillator strengths for 48 transitions with laser-induced plasmas (LIPs) produced by samples of varying Ni concentrations. To date however, no experimental transition probabilities for lines from higher-lying energy levels have been measured.

This investigation builds upon the subsequent works of \citet{Hartman2017} and \citet{Clear2026}. \citet{Hartman2017} extended experimental studies to the higher-lying levels of the {\Nidd} configuration for the first time by measuring lifetimes for seven such levels with TR-LIF. They also carried out pseudo-relativistic Hartree-Fock calculations, which were optimised with experimental energy levels from \citet{Shenstone1970}, to produce calculated oscillator strengths in the forms of both {\loggf} and \textsl{g\!A} for 477 transitions depopulating levels of the {\Nidd} configuration. To our knowledge, this is the only work published with experimental values for lifetimes of the {\Nidd} levels.

More recently, \citet{Clear2026} performed a large-scale calculation of theoretical transition probabilities and oscillator strengths for 118,000 {\NiII} electric-dipole (E1) transitions. They used the semi-empirical orthogonal operator method and fine-tuned the eigenvalues to the best available experimental energy levels from the extensive {\NiII} term analyses of \citet{Clear2022, Clear2023}. 

In this work, we present new branching fraction measurements of {\Nidd} -- {\Nidp} transitions  using high-resolution FT spectrometers at both Imperial College London (IC) and the National Institute of Standards and Technology (NIST). We report 174 transition probabilities and absolute oscillator strengths for transitions from 31 upper levels, normalised by the experimental level lifetimes of \citet{Hartman2017} where available, or otherwise the theoretically calculated lifetimes of \citet{Clear2026}. Of these 174 lines, 19 lines, found with experimental lifetimes, represent the first fully experimentally determined {\gf}-values for {\Nidd} -- {\Nidp} transitions. We compare our  results   with the theoretically calculated {\gf}-values of \citet{Clear2026}, \citet{Hartman2017}, and the extensive calculations of \citet{Kurucz2011}. 

\section{Experimental Procedure}
This work uses the branching fraction method of determining transition probabilities by combining experimental branching fractions for lines originating from the same upper energy level with experimental or theoretical level lifetimes to give  transition probabilities and absolute oscillator strengths. 

Branching fractions $\mathrm{BF}_{ul}$, for transitions from a common upper level $u$ to different lower levels, $l$, can be determined experimentally in an optically thin plasma according to:
\begin{equation}
\mathrm{BF}_{ul} = \frac{I_{ul}}{\sum_lI_{ul}} = \frac{c_{ul}I'_{ul}}{\sum_lc_{ul}I'_{ul}}
\label{eq: BF_exp}
\end{equation}
where $c_{ul}$ is the calibration factor due to the spectrometer response and $I'_{ul}$ is the observed line intensity \citep{Sikstrom2002}. Their product, the calibrated line intensity $I_{ul}$, is proportional to the transition probability of the line $A_{ul}$, giving \citep{Huber1986}:
\begin{equation}
\mathrm{BF}_{ul} = \frac{A_{ul}}{\sum_lA_{ul}}
\label{eq: BF_theo}
\end{equation}
Since the  sum of the transition probabilities  for all lines from a given upper level, $\sum_lA_{ul}$, is related to radiative lifetime $\tau_u$ of the  upper level by \citep{Huber1986}:
\begin{equation}
\tau_u = \frac{1}{\sum_lA_{ul}}
\label{eq: tau}
\end{equation}
the transition probability is obtained by:
\begin{equation}
A_{ul}=\frac{\mathrm{BF}_{ul}}{\tau_{u}}
\label{eq: A}
\end{equation}
which can be converted to the absolute {\gf}-value using
\begin{equation}
\log(\textsl{g}_l\!f) = \log[A_{ul}\textsl{g}_u\lambda^2\times 1.499 \times 10^{-14}]
\label{eq: loggf}
\end{equation}
where $f$ is the oscillator strength,  $\textsl{g}_l$ and  $\textsl{g}_u$ are the statistical weights of the lower and upper levels, respectively, and $\lambda$ is the transition wavelength in nanometers \citep{Thorne1999}.

\subsection{Spectrum Measurements}
As summarised in Table \ref{tab: spectra}, two sets of spectra were used to determine the branching fractions in this work.

\begin{deluxetable*}{ccDcccccc}
    \tablecaption{FT Spectra Used for Branching Fraction Measurements\label{tab: spectra}}
    \tablehead{
        Spectrum & Wavenumber Range & \multicolumn2c{Pressure} & Current & PMT & Filter & Resolution & Coadds & Spectrum Filename\\
         & Used ({\cm}) & \multicolumn2c{(mbar)}                & (mA) &  Detector       & & ({\cm}) & & 
    }
    \decimals
    \startdata
        A (IC) & 34000-52000 & 5.0 & 400 & R7154 & - & 0.05 & 20 & NiHe220614.050.071\\
        B (IC) & 34000-54000 & 7.5 & 750 & R7154 & - & 0.05 & 11 & NiHe220617.040.051\\
        C (IC) & 33000-54000 & 7.5 & 1000 & R7154 & - & 0.05 & 30 & NiHe220617.100.135\\
        D (IC) & 34500-48500 & 5.0 & 250 & R7154 & - & 0.05 & 35 & NiHe220629.010.044\\
        E (NIST) & 34000-52000 & 6.7 & 2000 & R7154 & - & 0.08 & 20 & Ni020723a.005\\
        F (NIST) & 24390-45000 & 9.9 & 2000 & R106UH & UG5 & 0.05 & 128 & Ni020923a.002\\
        G (NIST) & 25000-40000 & 10.1 & 2000 & R106UH & UG11 & 0.05 & 174 & Ni021323b.001\\
        H (NIST) & 34000-52000 & 10.0 & 2000 & R1220 & - & 0.08 & 256 & Ni021423a.001\\
    \enddata
\end{deluxetable*}

Spectra A--D were measured on the vacuum ultraviolet (VUV) FT Spectrometer (FTS) at IC \citep{Thorne1987, Thorne1996}, covering the spectral region between 31596 {\cm} and 63192 {\cm}. A high purity (99.99+\%, natural isotope composition) Ni cylinder, 40 mm in length and 8 mm in internal diameter, formed the cathode of a water-cooled hollow cathode lamp (HCL) \citep{Holmes2015}. The HCL was operated with He as the filler gas, at pressures of 5.0 or 7.5 mbar, with currents ranging from 250 to 1000 mA to ensure that strong lines are not affected by self-absorption. 

Spectra E--H were measured using the  FT700 VUV FTS \citep{Griesmann1999} at NIST, in the wavenumber range 22000 {\cm} to 55000 {\cm}. A 60 mm long high-purity (99.99+\%, natural isotope composition) Ni cylinder, of internal diameter of 8 mm, was mounted in a water-cooled HCL \citep{Danzmann1988}. The HCL was operated with He as the filler gas, at pressures between 6.7 and 10.1 mbar and a high current of 2A. 

Full details of the detectors, filters, and running conditions for each spectrum are specified in Table \ref{tab: spectra}. The resolution for each spectrum was chosen to fully resolve spectral lines to their Doppler widths. The signal-to-noise ratio (SNR) of the lines was improved by coadding multiple interferograms for each spectrum.

All spectra were intensity calibrated using deuterium (D\textsubscript{2}) standard lamp spectra. To verify that the spectrometer response was stable over the course of the {\NiII} experiment, D\textsubscript{2} lamp spectra were obtained before and after acquiring each Ni spectrum under identical experimental conditions. Only lines lying within the spectral region where the two D\textsubscript{2} spectra remained in good agreement within uncertainties were considered to be reliably calibrated, and retained in the subsequent analysis. The intensity calibrated wavenumber ranges for all spectra are listed in Table \ref{tab: spectra}. 

\begin{figure*}
    \includegraphics[width=\textwidth]{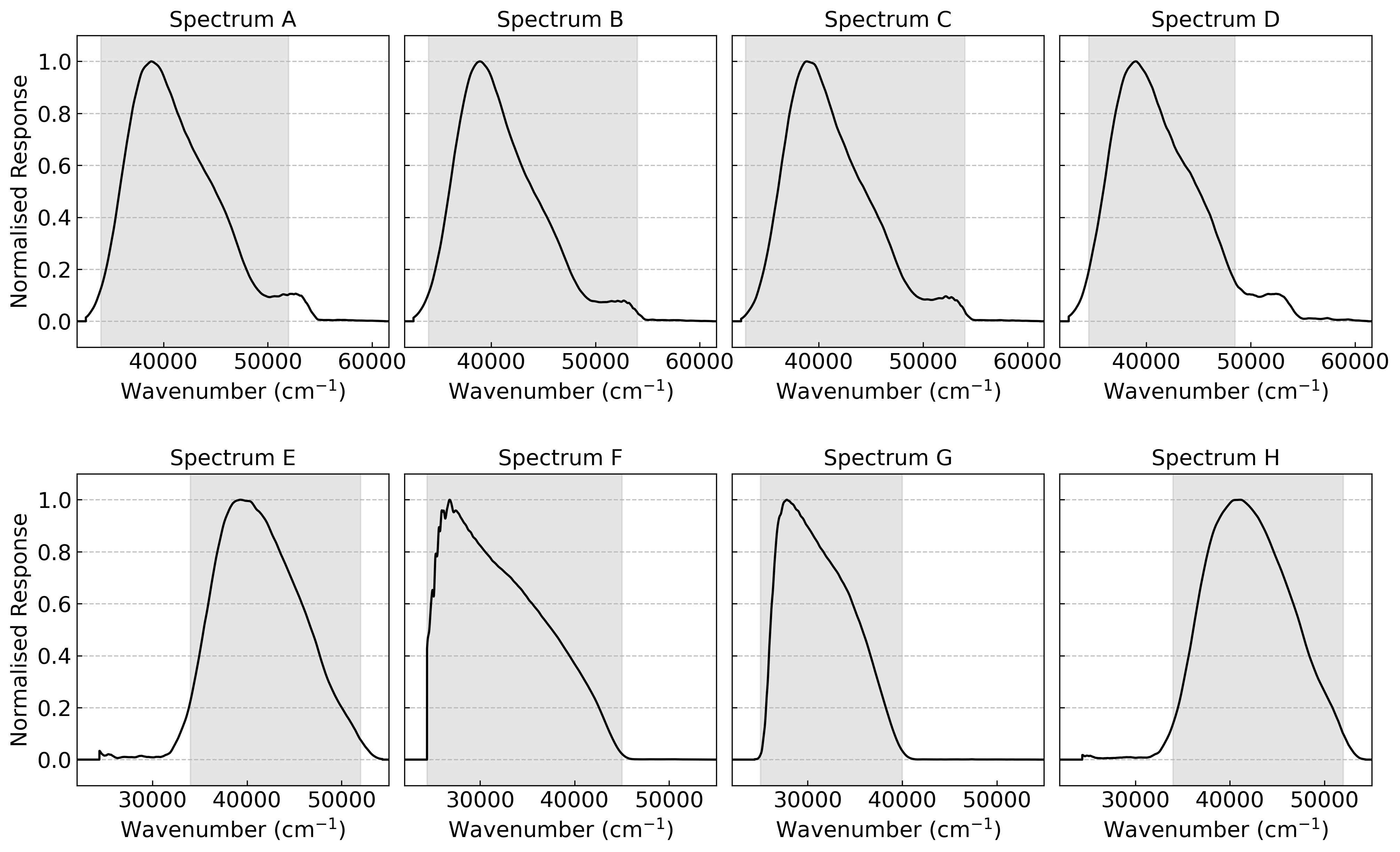}
    \caption{Normalised spectrometer response curves for the spectra measured in this work. The shaded regions indicate the spectral ranges over which the spectra could be accurately intensity calibrated and which were used in the subsequent determination of branching fractions.}
    \label{fig: response}
\end{figure*}

The D\textsubscript{2} standard lamps at IC and NIST were calibrated by the Physikalisch-Technische Bundesanstalt (PTB) in Germany, and “the relative extended uncertainty, which results from the relative standard uncertainty by multiplication with the extension factor k = 2" was reported in the calibration certificates provided. Here, we adopted “the relative standard deviation", corresponding to one standard deviation, as the uncertainty in the spectral radiance of the D\textsubscript{2} lamps. This uncertainty is 3.5\% at wavelengths longward of 172 nm for both lamps. The measured standard lamp spectra were filtered to match the resolution of their calibration certificate, and compared to the calibrated spectral radiances of the lamps to determine instrument response functions for each Ni spectrum. The resulting response functions are shown in Figure \ref{fig: response} for the eight spectra, A--H, and these were subsequently used to calibrate the relative line intensities.

\subsection{Branching Fraction Measurements}
To determine branching fractions, accurate relative line intensities are required, and these were obtained from the integrated area under the Voigt line profiles fitted with {\xgrem} \citep{Xgremlin} to each observed line. The peak height of a line was used, together with the spectrum noise level, to determine its SNR. Since the noise level for FT Spectroscopy is constant throughout a spectrum, it was estimated by averaging the  root-mean-square deviation from the mean intensity in five spectral regions containing only noise, i.e. regions without spectral lines. 

Following Equations (\ref{eq: BF_exp})--(\ref{eq: A}), we then used the computer program {\pybr} \citep{pybranch} to determine experimental branching fractions and transition probabilities from relative line intensities $I_{ul}$, calibrated using the response curves in Figure \ref{fig: response}, and level lifetimes $\tau_u$. A detailed description of the method used in {\pybr} to combine multiple spectra and evaluate uncertainties can be found in section 4.2 of \citet{Ward2023}. 

In general, the uncertainty of a line branching fraction is given by \citet{Ruffoni2013}, who derived the expression from \citet{Sikstrom2002}
\begin{equation}
(\frac{\Delta \mathrm{BF}_{ul}}{ \mathrm{BF}_{ul}})^2 = (1-2{\mathrm{BF}_{ul}})(\frac{\Delta I_{ul}}{I_{ul}})^2 + \sum_{j=1}^n (\mathrm{BF}_{uj})^2(\frac{\Delta I_{uj}}{I_{uj}})^2
\label{eq: BF unc}
\end{equation}
 where $\Delta I_{ul}$ is the uncertainty of the calibrated intensity of the transition from upper level $u$ to lower level $l$, and the subscript $j$ represents all possible connected lower levels. {\pybr} combines several sources of uncertainty to determine $\Delta I_{ul}$ following the procedure outlined in \citet{Sikstrom2002}, including those from line fitting, response functions, and, if necessary, additional scaling of spectra. 

The statistical error from fitting one line is given by
\begin{equation}
\frac{\Delta I'_{ul}}{I'_{ul}} = \alpha\frac{1}{\mathrm{SNR}\sqrt{n}}
\label{eq: I'_unc}
\end{equation}
where $n$ is the number of data points across the FHWM of the line and $\alpha$, an empirical constant from simulations, is $1.41 \pm 0.04$ for Gaussian and $1.60 \pm 0.04$ for Lorentzian lines \citep{Sikstrom2002}. For simplicity, {\pybr} uses a fixed intermediate value of 1.50 for all Voigt profiles \citep{Ward2023}.
 
The intensity calibration uncertainty from the spectrometer response functions was estimated by adding in quadrature the 3.5\% uncertainty in the spectral radiance of the standard lamp and the percentage variation observed in the D\textsubscript{2} lamp spectra. For the latter contribution, the maximum difference in response between pairs of D\textsubscript{2} spectra of 5.0\% was adopted, resulting in a total uncertainty of 6.1\%. It is worth noting that directly applying this uniform calibration uncertainty  to all lines within a spectrum results in an overestimation of the true uncertainty, as lines lying closer in wavenumber  will be less affected by the variation of response across the entire spectrum \citep{Sikstrom2002}. To account for this, {\pybr} was used to multiply the maximum uncertainty by a factor of $\Delta\sigma/W_\sigma$, where $\Delta\sigma$ is the wavenumber separation between a particular line of intensity $I_{ul}$ and the strongest line from the common upper level, and $W_\sigma$ is the total width of the calibrated Ni spectrum in which that line is observed \citep{pybranch}. Here, $W_\sigma$ corresponds to Column 2 of Table \ref{tab: spectra}.

\begin{deluxetable*}{cccDcccD}
    \tablecaption{Comparison of experimental and calculated lifetimes (ns) for {\NiII} {\Nidp} and {\Nidd} levels}
    \setlength{\tabcolsep}{10pt}
    \tablehead{
        \multicolumn3c{Level} &
        \multicolumn2c{Energy} &
        \multicolumn2c{Experiments} &
        \colhead{Calculation} &
        \multicolumn2c{Difference} \\
        \cline{1-3}\cline{6-7}
        \colhead{Configuration} &
        \colhead{Term} &
        \colhead{J} &
          \multicolumn2c{({\cm})} &
        \colhead{$\tau$ (ns)} &
        \colhead{Source} &
        \colhead{$\tau$ (ns)} &
        \multicolumn2c{(\%)}
    }
    \decimals
\startdata
$3d^8(^3F)4p$ & $^4D$ & 3.5 & 51557.7247 & $3.30 \pm 0.2$ & Fe99 & 3.42 & 3.5 \\
$3d^8(^3F)4p$ & $^4D$ & 2.5 & 52738.2988 & $3.25 \pm 0.2$ & Fe99 & 3.36 & 3.2 \\
$3d^8(^3F)4p$ & $^4G$ & 4.5 & 53365.0395 & $3.15 \pm 0.2$ & Fe99 & 3.21 & 1.8 \\
$3d^8(^3F)4p$ & $^4G$ & 5.5 & 53496.3843 & $2.80 \pm 0.2$ & Fe99 & 2.86 & 2.0 \\
$3d^8(^3F)4p$ & $^4D$ & 1.5 & 53634.5150 & $3.15 \pm 0.2$ & Fe99 & 3.34 & 6.2 \\
$3d^8(^3F)4p$ & $^4D$ & 0.5 & 54176.1448 & $3.20 \pm 0.2$ & Fe99 & 3.34 & 4.4 \\
$3d^8(^3F)4p$ & $^4G$ & 3.5 & 54262.5202 & $3.00 \pm 0.2$ & Fe99 & 3.04 & 1.4 \\
$3d^8(^3F)4p$ & $^4F$ & 4.5 & 54556.9460 & $3.00 \pm 0.2$ & Fe99 & 2.96 & -1.5 \\
$3d^8(^3F)4p$ & $^4G$ & 2.5 & 55018.5579 & $2.95 \pm 0.2$ & Fe99 & 2.96 & 0.5 \\
$3d^8(^3F)4p$ & $^2G$ & 4.5 & 55299.5972 & $3.00 \pm 0.2$ & Fe99 & 3.20 & 6.7 \\
$3d^8(^3F)4p$ & $^4F$ & 3.5 & 55417.6564 & $2.95 \pm 0.2$ & Fe99 & 2.85 & -3.5 \\
$3d^8(^3F)4p$ & $^4F$ & 2.5 & 56075.0648 & $2.90 \pm 0.2$ & Fe99 & 2.84 & -2.1 \\
$3d^8(^3F)4p$ & $^2G$ & 3.5 & 56371.2770 & $3.25 \pm 0.2$ & Fe99 & 3.50 & 7.7 \\
$3d^8(^3F)4p$ & $^4F$ & 1.5 & 56424.3010 & $3.00 \pm 0.2$ & Fe99 & 2.85 & -5.1 \\
$3d^8(^3F)4p$ & $^2F$ & 3.5 & 57080.3778 & $2.70 \pm 0.2$ & Fe99 & 2.78 & 2.9 \\
$3d^8(^3F)4p$ & $^2D$ & 2.5 & 57420.0091 & $2.00 \pm 0.2$ & Fe99 & 2.17 & 8.5 \\
$3d^8(^3F)4p$ & $^2F$ & 2.5 & 58493.0678 & $2.30 \pm 0.2$ & Fe99 & 2.40 & 4.2 \\
$3d^8(^3F)4p$ & $^2D$ & 1.5 & 58705.6944 & $2.00 \pm 0.2$ & Fe99 & 2.06 & 2.9 \\
$3d^8(^3F)4d$ & $^4D$ & 3.5 & 98467.093 & $1.28 \pm 0.1$ & Ha17 & 1.27 & -0.7 \\
$3d^8(^3F)4d$ & $^4H$ & 6.5 & 98822.419 & $1.25 \pm 0.1$ & Ha17 & 1.29 & 3.1 \\
$3d^8(^3F)4d$ & $^4G$ & 5.5 & 99132.589 & $1.32 \pm 0.1$ & Ha17 & 1.31 & -0.4 \\
$3d^8(^3F)4d$ & $^4F$ & 4.5 & 99154.623 & $1.20 \pm 0.1$ & Ha17 & 1.26 & 5.0 \\
$3d^8(^3F)4d$ & $^4D$ & 2.5 & 99559.165 & $1.37 \pm 0.1$ & Ha17 & 1.26 & -8.2 \\
$3d^8(^3F)4d$ & $^4H$ & 5.5 & 100309.129 & $1.30 \pm 0.1$ & Ha17 & 1.30 & 0.2 \\
$3d^8(^3F)4d$ & $^4G$ & 4.5 & 100619.096 & $1.25 \pm 0.1$ & Ha17 & 1.32 & 5.6 \\
\enddata
\tablecomments{Columns 1--4 give experimental energy levels and designations taken from \citet{Clear2022}; Column 5 gives the experimental lifetimes of Fe99 — \citet{Fedchak1999} and Ha17 — \citet{Hartman2017} with uncertainties; Column 7 gives the theoretically calculated lifetimes of \citet{Clear2026}; Column 8 gives percentage differences between experimental and calculated lifetimes.}
    \label{tab: tau}
\end{deluxetable*}

 A majority of the lines appear in multiple spectra. To place multiple measurements of a line on a common intensity scale, a reference line, usually the strongest transition from the upper level of interest, was selected in {\pybr} and a scaling factor applied to each spectrum to result in the same intensity value for the reference line in each spectrum. In the case of some upper levels investigated, spectrum F or G, whose responses peak at lower wavenumbers, did not contain the selected reference line   but did provide additional lines not observed in other spectra. Therefore, ideally, a transfer line from the upper level of interest, observed in both the spectrum to be scaled and at least one of the other spectra containing the reference line, was chosen to determine the normalisation factor \citep{Pickering2001FeII, Pickering2001TiII}.  If no such line was available, we instead used the intensity ratio of a line from another  upper {\Nidd} level, of similar energy to the target upper level, to provide the scaling transfer factor in the overlapping region. We then increased the uncertainty of all lines in the spectrum without the reference line to account for this scaling \citep{Ward2023}.
 
The calibration uncertainty $\Delta c_{ul}$ (with the scaling uncertainty where applicable) as described above was combined in quadrature with the statistical uncertainty $\Delta I'_{ul}$ from the line SNR in Equation (\ref{eq: I'_unc}) to obtain the total uncertainty in the relative intensity of each individual line
\begin{equation}
(\frac{\Delta I_{ul}}{I_{ul}})^2 = (\frac{\Delta I'_{ul}}{I'_{ul}})^2 + (\frac{\Delta c_{ul}}{c_{ul}})^2
\label{eq: I_unc}
\end{equation}
The resulting total uncertainty $\Delta I_{ul}$ was subject to a minimum of 4.6\% to prevent potential systematic errors in the calibration from disproportionately influencing the results \citep{Ward2023}.

Instead of relying on $I_{ul}$ and $\Delta I_{ul}$ from a single spectrum, {\pybr} computes the average relative line intensity for each transition from a chosen upper level across all spectra, with each measurement weighted by the inverse square of its calibrated intensity uncertainty. The resulting weighted average line intensity and its associated uncertainty were then used for $I_{ul}$ and $\Delta I_{ul}$ in Equation (\ref{eq: BF unc}). For strong transitions, lines from each spectrum were weighted similarly due to the minimum uncertainty limit. For weak transitions, spectrum H was always favoured during this process due to its exceptionally high SNR from 256 coadds, and lines from this spectrum dominated the final weighted line intensity. Agreement between measurements from different spectra, within their uncertainties, provided independent confirmation of the reliability of our results, in addition to ruling out any possible issue with self absorption of very strong lines.

Ideally, a complete set of all possible transitions to lower levels from an upper level is observed. However, where this is not possible, an unobserved fraction remains, known as the residual branching fraction. This residual BF, due to missing lines which were either too weak to be observed or outside the available spectral range, was determined by {\pybr} using the calculated transition probabilities of \citet{Clear2026}, and used to correct for the sum of $I_{ul}$ in Equation (\ref{eq: BF_exp}).

\subsection{Determination of transition probabilities and {\loggf}s}
Once the BF  and their uncertainties, $\Delta\mathrm{BF}_{ul}$, had been found, we proceeded to determine the line transition probabilities and {\gf}-values.
The uncertainty of the $\mathrm{BF}_{ul}$ can be propagated to $A_{ul}$ by
\begin{equation}
(\frac{\Delta A_{ul}}{A_{ul}})^2 = (\frac{\Delta\mathrm{BF}_{ul}}{\mathrm{BF}_{ul}})^2  + (\frac{\Delta\tau_u}{\tau_u})^2 
\label{eq: A unc}
\end{equation}
where $\Delta\tau_u$ is the uncertainty of the upper level lifetime. For the seven {\Nidd} levels with experimental lifetimes in \citet{Hartman2017}, we adopted the experimental values of $\tau_u$ and associated uncertainties. For the remaining {\Nidd} levels, we derived theoretical lifetimes according to Equation (\ref{eq: tau}) using the calculated transition probabilities of \citet{Clear2026}. To estimate the reliability of these theoretical lifetimes, we compared them with the seven {\Nidd} experimental lifetimes of \citet{Hartman2017} and the 18 experimental lifetimes of {\Nidp} levels of \citet{Fedchak1999}. As shown in Table \ref{tab: tau}, our theoretical lifetimes were found to agree with the experimental lifetimes within approximately 8\%. This value, 8\%,  was therefore used as the relative uncertainty for all theoretically calculated lifetimes in this work. Finally, the uncertainty in $\log(\textsl{g}_lf)$ follows as
\begin{equation}
\Delta\log(\textsl{g}_l\!f)=\log(1+ \frac{\Delta A_{ul}}{A_{ul}})
\end{equation}

\section{Results}

\begin{figure*}
        \includegraphics[width=\textwidth]{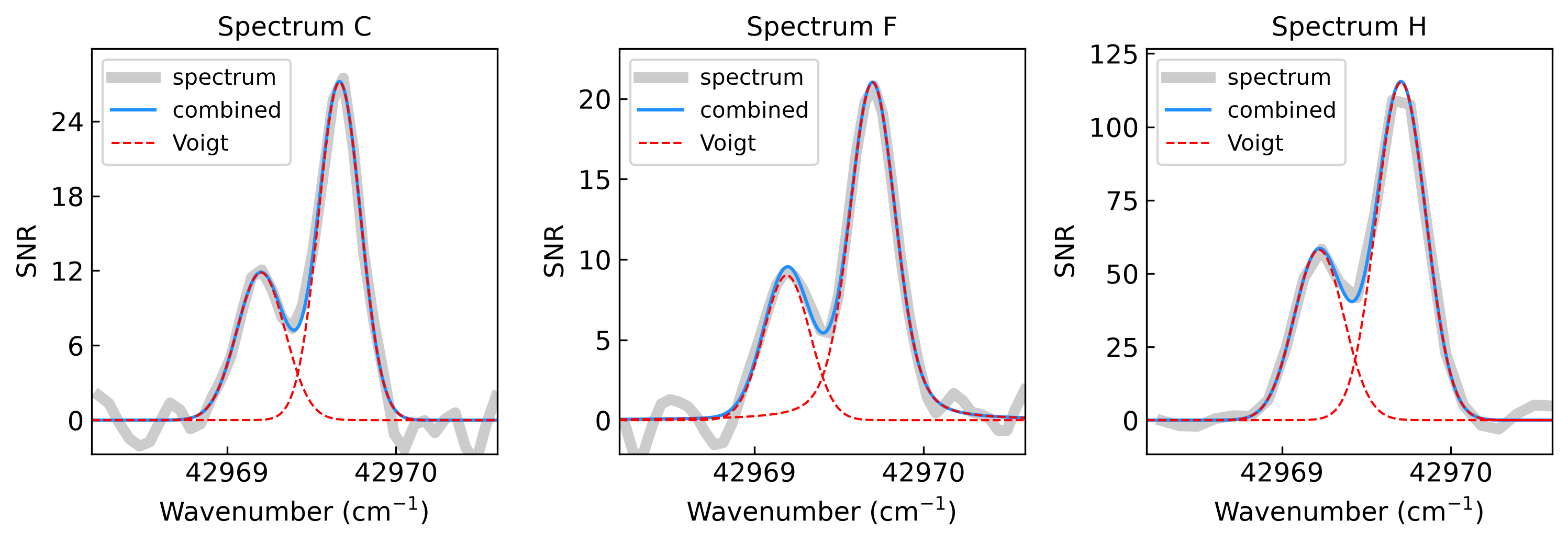}
    \caption{An example of the fitting of components of a blend of two lines observed in three spectra: Spectrum  C (left), Spectrum F (middle) and Spectrum G (right).  The  lines of transitions $3d^8(^3F) 4d \;^2F_{7/2}$ - $ 3d^8 (^3F)4p \;^2G_{7/2}$  at  42969.0779 {\cm} and $3d^8 (^3F)4d \;^4F_{5/2}$ - $ 3d^8(^3F) 4p \;^2D_{5/2}$  at 42969.3216 {\cm} are blended with each other and fitted simultaneously. In each plot, the observed spectrum is shown in grey, the combined fit in blue, and the individual fitted Voigt profiles for the two lines as red dashed curves.}
    \label{fig: blend}
\end{figure*}

We consider an upper level to be sufficiently complete, and the associated oscillator strengths to be reliable, when more than 95\% of all possible transitions to lower levels were observed, i.e. the residual branching fraction was $<5\%$. The fully experimental and semi-empirical transition probabilities of this work are given in Tables \ref{tab: loggf_exp} and \ref{tab: loggf_theo} respectively. Table \ref{tab: loggf_exp} presents entirely experimental transition probabilities and {\loggf}s for lines from four levels  determined using experimental BFs of our work and the measured lifetimes of \citet{Hartman2017}, with their reported uncertainties. Table \ref{tab: loggf_theo} gives transition probabilities  (and {\loggf}s) for lines from 27 upper levels for which the transition probabilities were normalised by theoretical lifetimes derived from the calculated transition probabilities of \citet{Clear2026}, with a fixed lifetime uncertainty of 8\%, as estimated in the previous section.  In addition, for completeness, we also include levels for which only a single strong transition with a branching fraction close to unity was observed. In these cases, the transition probability is essentially the reciprocal of the lifetime, with a small correction due to weak residuals. These values are derived entirely from the works of \citet{Clear2026} and \citet{Hartman2017} and do not constitute an independent determination of oscillator strengths. Nevertheless, our measurement of one dominant line after an extensive search for other predicted transitions provides experimental confirmation that the remaining transitions are indeed as weak as predicted by theory.

Table \ref{tab: loggf_res} lists four additional levels which are less than 95\% complete. The associated branching fractions should therefore be treated as branching ratios, and the resulting oscillator strengths are reliable only on a relative scale. 

The remaining {\NiII} {\Nidd} levels that are not presented here either have a substantial fraction of their transitions lying outside the available wavenumber range of this work or contain potential line blends that introduce inconsistencies between spectra and require further investigation. Analysis of these levels is ongoing.

For all three tables discussed above, the sets of transitions are arranged in order of increasing upper level energy,  and, within each set from a particular  upper level, in order of decreasing branching fraction. Each group of transitions is preceded by a header specifying the upper level energy, level label, lifetime with uncertainty, and completeness of the total branching fraction. Each transition is presented with the energy and J-value of the lower level, Ritz wavenumber, branching fraction with percentage uncertainty, transition probability with percentage uncertainty, and {\loggf} with uncertainty in dex. The energy levels, level labels, and Ritz wavenumbers presented in the tables were all taken from \citet{Clear2022}. The calculated {\loggf} values from \cite{Clear2026}, \citet{Hartman2017} and \citet{Kurucz2011} are also listed for comparison. \citet{Hartman2017} only published predictions for energy levels up to 103,663.342 {\cm}, so the corresponding column is left blank for levels beyond this in Table \ref{tab: loggf_theo} and is  fully omitted from Table \ref{tab: loggf_res}.  We  include the theoretical calculations of {\loggf}s of \citet{Kurucz2011} in our comparisons, as the late Bob Kurucz's atomic data have been used extensively in stellar model atmosphere codes for many years.

During the analysis, the uncertainties of particular lines needed to be increased due to issues with the spectral lines. These lines are annotated with tags in the tables. The tag "B" indicates a known partial blend with another spectral line. If multiple measurements were available for such a line, the maximum deviation from the weighted average was taken as the uncertainty of the relative line intensity $\Delta I_{ul}$. However, if the transition was observed in only one spectrum, we arbitrarily increased $\Delta I_{ul}$ by a factor of 2. For example, as shown in Figure \ref{fig: blend}, two blended {\NiII} {\Nidd} -- {\Nidp} lines at 42969.0779 {\cm} and 42969.3216 {\cm} are plotted as seen in spectra C, F and H. These blended lines were simultaneously fitted with Voigt profiles in {\xgrem}, and it is apparent from the plots that the blend caused the total intensity to be distributed slightly differently between the two lines across spectra.
In another example, two lines at 41683.6362 {\cm} and 43070.7239 {\cm} are marked with “U" and their uncertainties were increased due to unexplained scatter in the measured line intensities between spectra. There is a possibility of blending, although we found no obvious evidence in the line shape or the calculations. The second line was also reported as a blend with an unidentified line by \citet{Clear2022} during the optimisation of {\NiII} energy levels, again suggesting a potential issue.

The full linelist, sorted by increasing wavenumber, is available in machine-readable format in the online material (see Table \ref{tab: ll} for an extract). The table provides, for each transition, the  Ritz wavenumber, wavelength in air and vacuum, the energies and J-values of the upper and the lower levels, experimental transition probability with percentage uncertainty, experimental {\loggf} value with uncertainty in dex, and theoretical {\loggf} values from \citet{Clear2026} and \citet{Kurucz2011}.

\begin{figure}
    \includegraphics[width=\columnwidth]{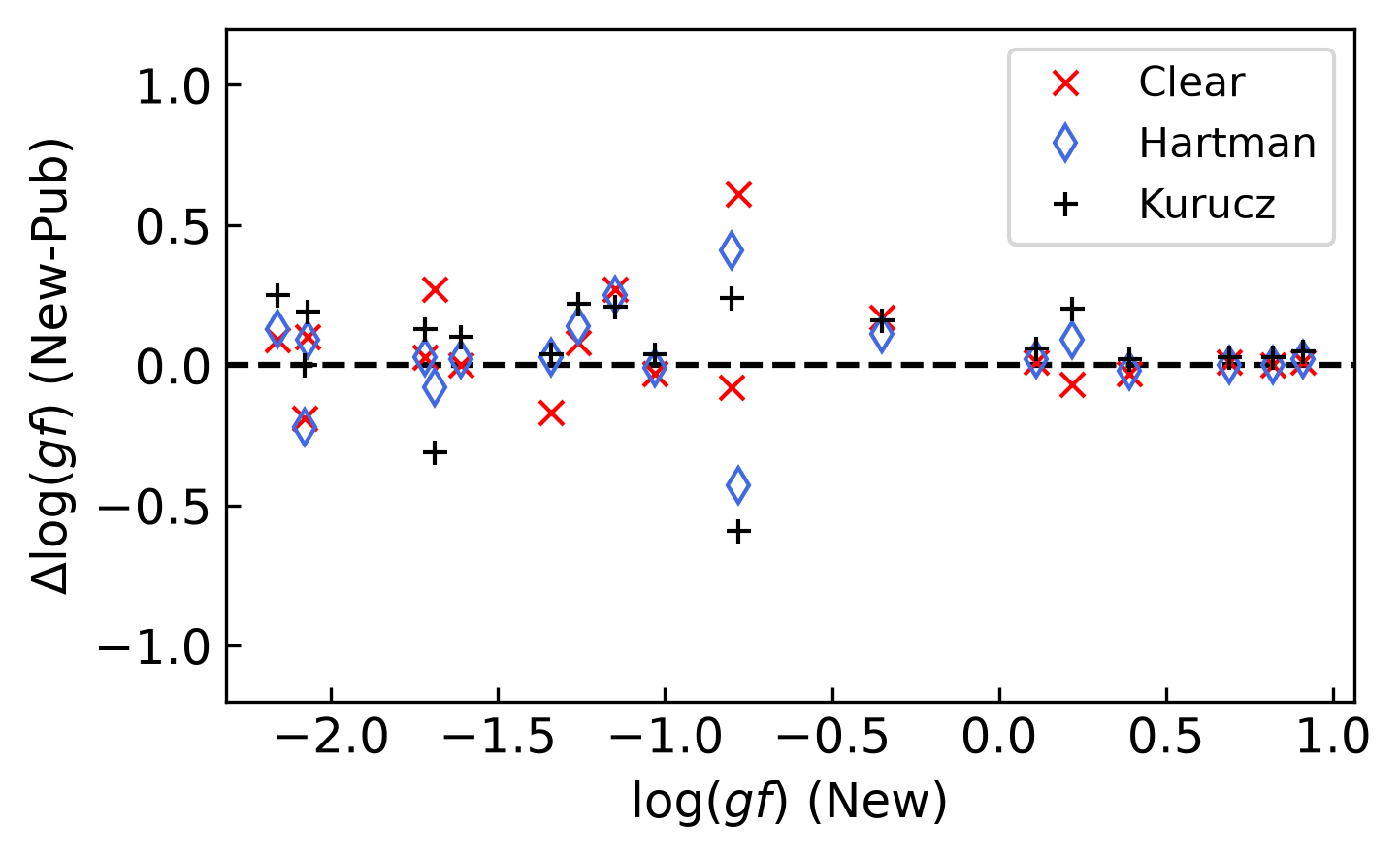}
    \caption{Comparisons of fully experimental {\loggf} values from this work, normalised by the experimental lifetimes of \citet{Hartman2017}, and previous theoretical calculations of \citet{Clear2026} (red), \citet{Hartman2017} (blue), and \citet{Kurucz2011} (black).}
    \label{fig: loggf_exp}
\end{figure}

We compare our new {\loggf} values with three sets of previous theoretical calculations by \citet{Clear2026}, \citet{Hartman2017} and \citet{Kurucz2011}. Figure \ref{fig: loggf_exp} contains the 19 transitions with our purely experimental {\loggf} values from Table \ref{tab: loggf_exp}. In general, the theoretical {\loggf}s   are consistent with our experimental {\loggf}s, with differences within \textpm 0.4 dex. Of the four {\Nidd} upper levels involved, the levels at 98467.093 {\cm} and 98822.419 {\cm} are predicted to be very pure by \citet{Clear2026}, while the other two, at 99132.589 {\cm} and 100309.129 {\cm}, appear relatively mixed, with leading eigenvector percentages of 78\% and 54\% respectively. This mixing likely explains the larger discrepancies between the new and previously published oscillator strengths for the two lines at 45767.5497 {\cm} and 46944.0894 {\cm} with experimental {\loggf} values of approximately -0.8, as calculations are known to be less accurate in regions of strong configuration mixing \citep{Clear2026}. 

\begin{figure}
\centering
\gridline{\fig{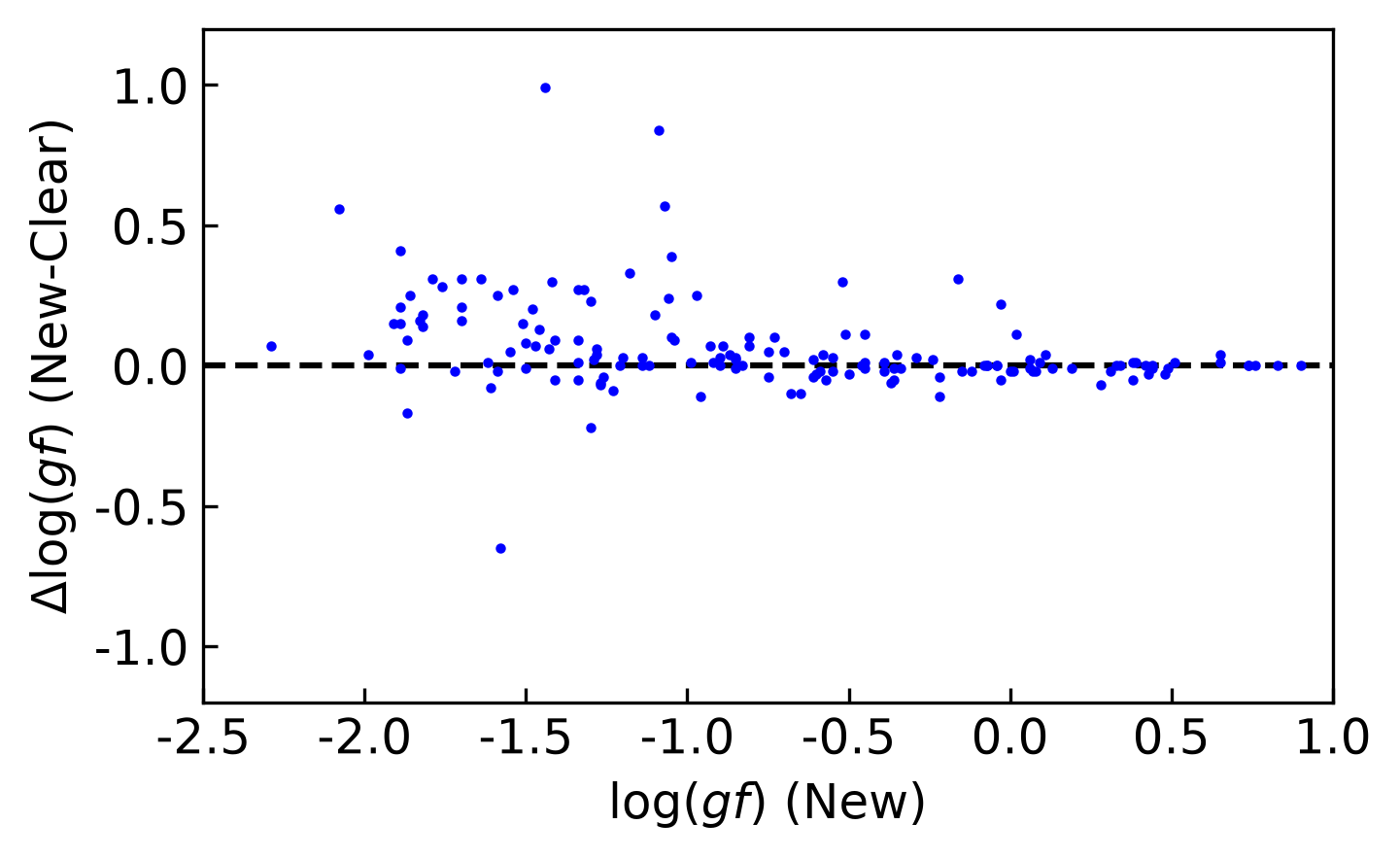}{\columnwidth}{(a)}}
\vspace{-10pt}
\gridline{\fig{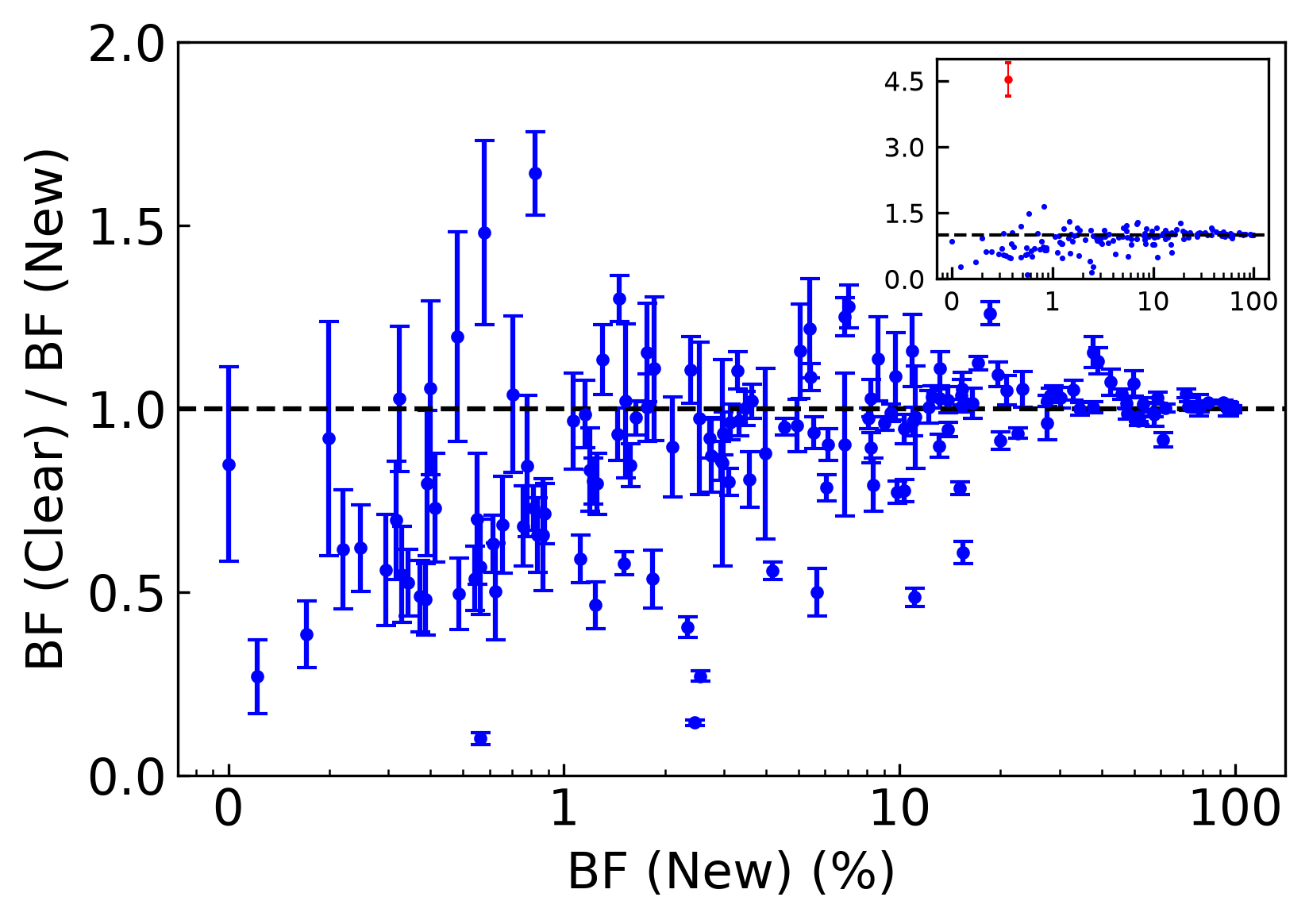}{0.95\columnwidth}{(b)}}
    \caption{Panel (a): Comparison of {\loggf} values from this work, normalised by the theoretical lifetimes of \citet{Clear2026}, and the theoretically calculated {\loggf}s of \citet{Clear2026}.
         Panel (b): Comparison of BFs from this work and \citet{Clear2026}. The error bars only include the uncertainty in experimental BFs.}
    \label{fig: loggf_theo_Clear}
\end{figure}

\begin{figure}
    \includegraphics[width=\columnwidth]{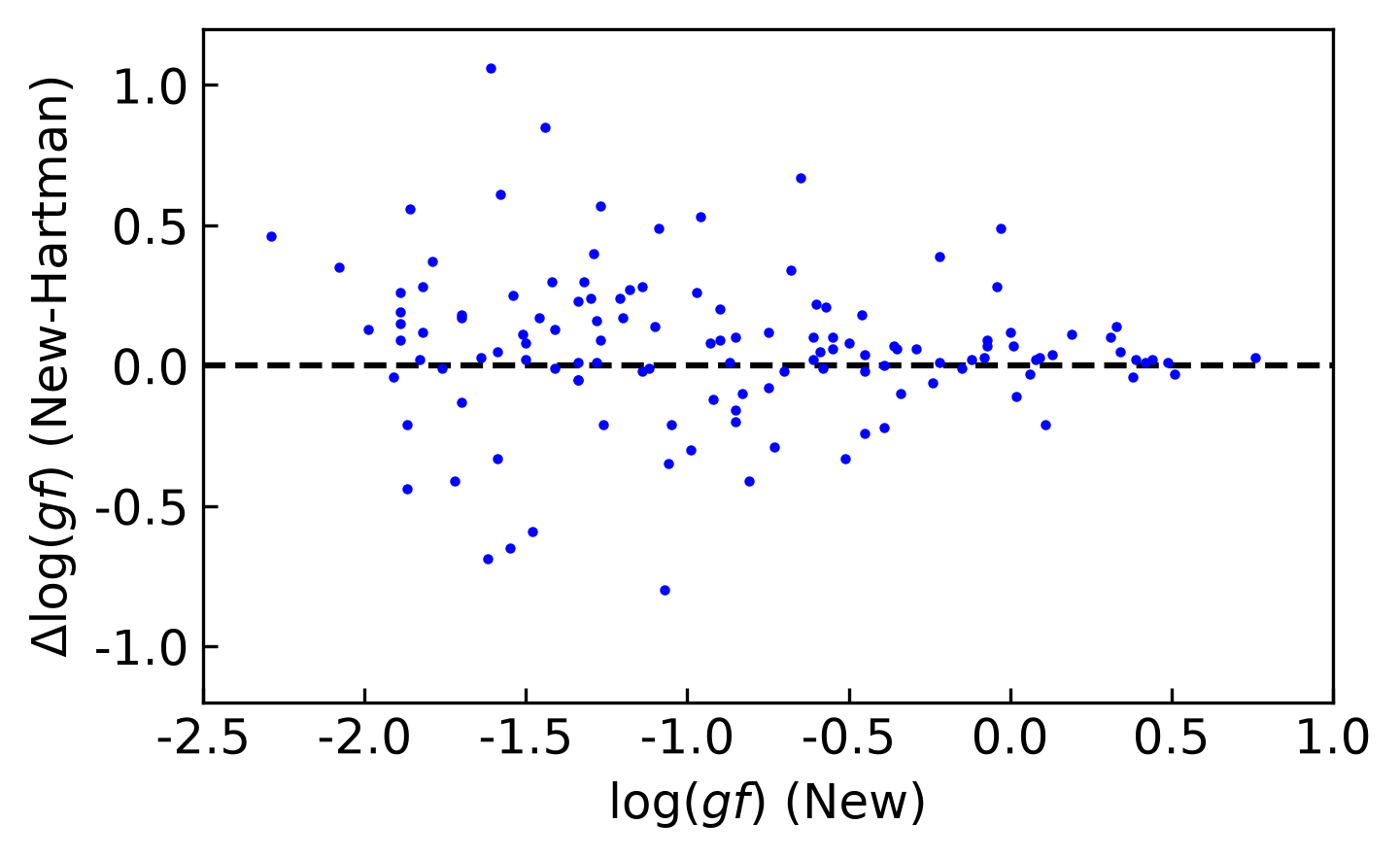}
    \caption{Comparison of {\loggf} values from this work, normalised by the theoretical lifetimes of \citet{Clear2026}, and the theoretically calculated {\loggf}s of \citet{Hartman2017}.}
    \label{fig: loggf_theo_Hartman}
\end{figure}

\begin{figure}
    \includegraphics[width=\columnwidth]{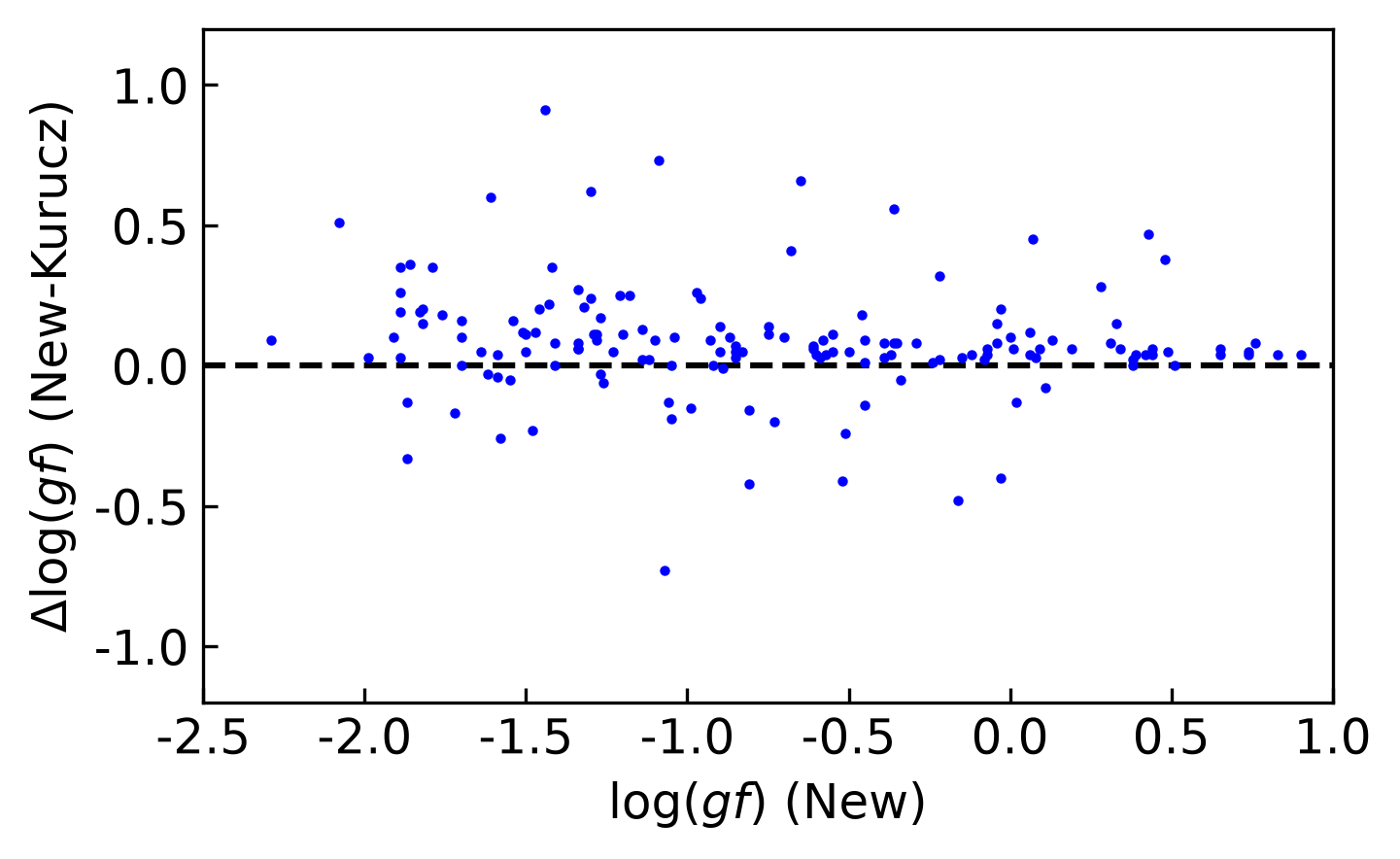}
    \caption{Comparison of {\
    loggf} values from this work, normalised by the theoretical lifetimes of \citet{Clear2026}, and the theoretically calculated {\loggf}s of \citet{Kurucz2011}.}
    \label{fig: loggf_theo_Kurucz}
\end{figure}

Figures \ref{fig: loggf_theo_Clear}, \ref{fig: loggf_theo_Hartman} and \ref{fig: loggf_theo_Kurucz} compare the 155 transitions with partially experimental {\loggf}s of Table \ref{tab: loggf_theo} with previously published values. As shown in Figure \ref{fig: loggf_theo_Clear}, the {\loggf}s
of \citet{Clear2026}  show good agreement with our new {\loggf} values down to approximately $-1.0$. For transitions with smaller oscillator strengths, the scatter increases due to the limitations of accuracy in calculations for weak lines. The comparisons with the theoretical {\loggf}s of \citet{Hartman2017} and \citet{Kurucz2011}, shown in Figures \ref{fig: loggf_theo_Hartman} and \ref{fig: loggf_theo_Kurucz}, illustrate similar trends but with greater scatter overall. 

All three figures are set to the same scale to allow direct visual comparison.
This is expected partially because \citet{Clear2026} fitted their theoretically calculated energy levels to the most accurate experimental energy levels to date from \citet{Clear2022, Clear2023}, whilst \citet{Hartman2017} and \citet{Kurucz2011} used the less accurate experimental values of \citet{Shenstone1970}. However, it is more likely that the agreement between observed and calculated {\loggf}s is probably overestimated in Figure \ref{fig: loggf_theo_Clear}a because the lifetimes used for the {\loggf}s were found from the very calculations that are in the comparison.  As these lifetimes are dominated by the stronger lines, the agreement for these would appear better than expected.   Figure \ref{fig: loggf_theo_Clear}b  shows the comparison of BFs from this work and those calculated using \citet{Clear2026}.  It is clear that theoretically calculated BFs are no substitute for experimental BFs, particularly for weaker transitions.
 One outlier {\loggf} from \citet{Hartman2017}, for the line at 43960.656 {\cm}, lies beyond the limits of the displayed $y$-axis range in Figure \ref{fig: loggf_theo_Hartman}. This transition has a cancellation factor (CF) of 0.00, and according to \citet{Hartman2017}, transitions with a CF below 0.05 should be treated with caution due to cancellation effects.

\setlength{\floatsep}{15pt}

\FloatBarrier
\section{Summary}
We have measured {\NiII} transitions, depopulating high-lying {\Nidd} levels, in the UV region with high-resolution FT spectroscopy, and determined absolute oscillator strengths and transition probabilities  for 174 lines from 31 upper levels between 98561.062 {\cm} and 122847.362 {\cm}. Branching fractions for 19 lines from four upper levels were combined with measured lifetimes of \citet{Hartman2017} to give fully experimentally determined {\loggf} values for the first time. For the remaining lines, oscillator strengths were placed on an absolute scale using theoretical lifetimes derived from the calculated transition probabilities of \cite{Clear2026}. Good agreement is found between our measurements and the calculated {\loggf} values of \citet{Clear2026} for strong lines with {\loggf} values above -1.0, while the weaker transitions show larger scatter. The calculations of \cite{Clear2026}, with eigenvalues fine-tuned to the most accurate experimental {\NiII} energy levels of \citet{Clear2022, Clear2023}, reproduce our new oscillator strengths more closely than the theoretical calculations of \citet{Hartman2017} and \cite{Kurucz2011}. This work represents the first experimentally determined branching fractions, transition probabilities and oscillator strengths for {\NiII}  {\Nidd} -- {\Nidp} transitions and will enable accurate chemical abundance determinations using these transitions in astrophysical spectra.

\begin{acknowledgments}
CPC and JCP thank the STFC of the UK for their support through the grants ST/W000989/1 and UKRI1188.
\end{acknowledgments}

\begin{deluxetable*}{DDDcDDDDDDDD}[htb!]
    \tablecaption{Fully experimental transition probabilities and {\loggf}s of {\NiII}, determined by combining experimental branching fractions and the experimentally measured lifetimes of \citet{Hartman2017}}
    \setlength{\tabcolsep}{4pt}
    \tablehead{
        \multicolumn4c{Lower Level} & \multicolumn2c{$\sigma$} & \colhead{Tag} & \multicolumn2r{BF} & \multicolumn2c{$u_r$(BF)} & \multicolumn2c{$A$} & \multicolumn2c{$u_r(A)$} & \multicolumn8c{{\loggf}}\\
        \cline{1-4}\cline{16-23}
        \multicolumn2c{E ({\cm})} & \multicolumn2c{J} & \multicolumn2c{({\cm})} & & \multicolumn2r{(\%)} & \multicolumn2c{(\%)} & \multicolumn2r{(10$^6$ s$^{-1}$)} & \multicolumn2c{(\%)} & \multicolumn2c{this work} & \multicolumn2c{Clear} & \multicolumn2c{Hartman} & \multicolumn2c{Kurucz}
    }
    \decimals
\startdata
\enddata
    \tablecomments{Columns are as follows: E and J, lower energy level of the transition and its J-value from \citet{Clear2022}; $\sigma$, Ritz wavenumber from \citet{Clear2022}; BF and $u_r$(BF), branching fraction and its relative uncertainty; $A$ and $u_r(A)$, absolute transition probability in 10$^6$ s$^{-1}$ and its relative uncertainty; absolute {\loggf} from this work and its uncertainty in dex, and theoretical {\loggf}s from \citet{Clear2026}, \citet{Hartman2017} and \citet{Kurucz2011} for comparison. The tag “B" indicates a partial blend.}
    \label{tab: loggf_exp}
\end{deluxetable*}

\clearpage
\startlongtable
\begin{deluxetable*}{DDDcDDDDDDDD}
    \tablecaption{Absolute transition probabilities and {\loggf}s, determined by combination of experimental branching fractions and the theoretical lifetimes of \citet{Clear2026}}
    \setlength{\tabcolsep}{4pt}
    \tablehead{
        \multicolumn4c{Lower Level} & \multicolumn2c{$\sigma$} & \colhead{Tag} & \multicolumn2r{BF} & \multicolumn2c{$u_r$(BF)} & \multicolumn2c{$A$} & \multicolumn2c{$u_r(A)$} & \multicolumn8c{{\loggf}}\\
        \cline{1-4}\cline{16-23}
        \multicolumn2c{E ({\cm})} & \multicolumn2c{J} & \multicolumn2c{({\cm})} & & \multicolumn2r{(\%)} & \multicolumn2c{(\%)} & \multicolumn2r{(10$^6$ s$^{-1}$)} & \multicolumn2c{(\%)} & \multicolumn2c{this work} & \multicolumn2c{Clear} & \multicolumn2c{Hartman} & \multicolumn2c{Kurucz}
    }
    \decimals
\startdata
\enddata
 \tablecomments{Columns are as follows: E and J, lower energy level of the transition and its J-value from \citet{Clear2022}; $\sigma$, Ritz wavenumber from \citet{Clear2022}; BF and $u_r$(BF), branching fraction and its relative uncertainty; $A$ and $u_r(A)$, absolute transition probability in 10$^6$ s$^{-1}$ and its relative uncertainty; absolute {\loggf}  from this work and its uncertainty in dex, and theoretical {\loggf}s from \citet{Clear2026}, \citet{Hartman2017} and \citet{Kurucz2011} for comparison. The tag “B" indicates a partial blend and the tag “U" indicates an unresolved issue.}
    \label{tab: loggf_theo}
\end{deluxetable*}

\begin{deluxetable*}{DDDcDDDDDDDD}
    \tablecaption{Approximate absolute transition probabilities and {\loggf}s for {\NiII} transitions from upper levels that are less than 95\% complete, determined by combining experimental branching fractions and the theoretical lifetimes of \citet{Clear2026}}
    \setlength{\tabcolsep}{3pt}
    \tablehead{
        \multicolumn4c{Lower Level} & \multicolumn2c{$\sigma$} & \colhead{Tag} & \multicolumn2r{BF} & \multicolumn2c{$u_r$(BF)} & \multicolumn2c{$A$} & \multicolumn2c{$u_r(A)$} & \multicolumn6c{{\loggf}}\\
        \cline{1-4}\cline{16-21}
        \multicolumn2c{E ({\cm})} & \multicolumn2c{J} & \multicolumn2c{({\cm})} & & \multicolumn2r{(\%)} & \multicolumn2c{(\%)} & \multicolumn2r{(10$^6$ s$^{-1}$)} & \multicolumn2c{(\%)} & \multicolumn2c{this work} & \multicolumn2c{Clear} & \multicolumn2c{Kurucz}
    }
    \decimals
\startdata
\enddata
     \tablecomments{Columns are as follows: E and J, lower energy level of the transition and its J-value from \citet{Clear2022}; $\sigma$, Ritz wavenumber from \citet{Clear2022}; BF and $u_r$(BF), branching fraction and its relative uncertainty; $A$ and $u_r(A)$, absolute transition probability in 10$^6$ s$^{-1}$ and its relative uncertainty; absolute {\loggf} from this work and its uncertainty in dex, and theoretical {\loggf}s from \citet{Clear2026}, \citet{Hartman2017} and \citet{Kurucz2011} for comparison. The tag “B" indicates a partial blend.}
    \label{tab: loggf_res}
\end{deluxetable*}

\begin{deluxetable*}{DDDDDDDDDDDDDc}
    \tablecaption{{\NiII} lines with new experimental transition probabilities and {\loggf}s}
    \setlength{\tabcolsep}{4pt}
    \tablehead{
        \multicolumn2c{$\sigma$} & \multicolumn2c{$\lambda_{air}$} & \multicolumn2c{$\lambda_{vac}$} & \multicolumn2c{E$_u$} & \multicolumn2c{J$_u$} & \multicolumn2c{E$_l$} & \multicolumn2c{J$_l$} & \multicolumn2c{$A$} & \multicolumn2c{Unc.} & \multicolumn8c{{\loggf}} & Tag\\
        \cline{19-26}
        \multicolumn2c{({\cm})} & \multicolumn2c{(nm)} & \multicolumn2c{(nm)} & \multicolumn2c{({\cm})} & \multicolumn2c{} & \multicolumn2c{({\cm})} & \multicolumn2c{} & \multicolumn2c{(10$^6$ s$^{-1}$)} & \multicolumn2c{(\%)} & \multicolumn2c{this work} & \multicolumn2c{Unc.} & \multicolumn2c{Clear} & \multicolumn2c{Kurucz} & }
     \decimals
\startdata
\enddata
    \tablecomments{Columns 1--7 give the Ritz wavenumber, wavelength in air, wavelength in vacuum, upper energy level with its J-value, and lower energy level with its J-value from \citet{Clear2022}; Columns 8--9 give the transition probability from this work in units of $10^6$ s$^{-1}$ and its relative uncertainty; Columns 10--11 give the absolute {\loggf} from this work and its uncertainty in dex; Columns 12--13 give the theoretical {\loggf}s from \citet{Clear2026} and \citet{Kurucz2011} for comparison; Column 14 gives a tag for an observed transition, with “B"  indicating a partial blend and “U"  indicating an unresolved issue. The complete table is available in machine-readable format in the online material. A portion of this table is reproduced here to illustrate its form and content.}
    \label{tab: ll}
\end{deluxetable*}






\FloatBarrier
\bibliography{Ni_TP_main}
\bibliographystyle{aasjournalv7.1}



\end{document}